\documentclass[conference]{IEEEtran}

\usepackage{graphicx}
\usepackage[caption=false,font=footnotesize]{subfig}

\usepackage{stfloats}
\usepackage{array}

\usepackage{algpseudocode}

\usepackage{verbatim}

\usepackage{tikz}
\usepackage{multirow}

\usepackage{mdwmath}
\usepackage{mdwtab}
\usepackage{amsmath}
\usepackage{amssymb}

\usepackage[utf8x]{inputenc}

\usepackage[printonlyused,smaller]{acronym}

\title{CPU Temperature and Power Modeling}
\author{Karel De Vogeleer}

\newcommand{\BSIM}{{\small BSIM}~}

\newcommand{\Pcpu}{P_\mathrm{\smaller cpu}}
\newcommand{\Pdyn}{P_\mathrm{dyn}}
\newcommand{\Psh}{P_\mathrm{short}}
\newcommand{\Pleak}{P_\mathrm{leak}}
\newcommand{\Pl}{P_{\mathrm{leak}}}

\newcommand{\dgr}{{$^\circ$C}}
\newcommand{\erf}{\mathrm{erf}}

\acrodef{AAC}{Advanced Audio Coding}
\acrodef{ADC}{Analog Digital Converter}
\acrodef{AGU}{Address Generating Unit}
\acrodef{ALU}{Algorithmic Logical Unit}
\acrodef{AVM}{Astute Virtual Machine}
\acrodef{ASIC}{Application-Specific Integrated Circuit}
\acrodef{BEEBS}{Bristol Energy Efficiency Benchmark Suite}
\acrodef{BSIM}{Berkeley Short-channel IGFET Model}
\acrodef{BTBT}{band-to-band tunneling}
\acrodef{CISC}{Complex Instruction Set Computer}
\acrodef{CMOS}{complementary metal-oxide-semiconductor}
\acrodef{CPU}{Central Processing Unit}
\acrodef{DFG}{Data Flow Graph}
\acrodef{DDIO}{Data Direct I/O}
\acrodef{DLP}{Data Level Parallelism}
\acrodef{DMA}{Direct Memory Access}
\acrodef{DRAM}{Dynamic Random-Access Memory}
\acrodef{DSP}{Digital Signal Processor}
\acrodef{DVS}{Dynamic Voltage Scaling}
\acrodef{DVFS}{Dynamic Voltage and Frequency Scaling}
\acrodef{DPM}{Dynamic Power Management}
\acrodef{EDD}{Energy-Delay Diagram}
\acrodef{EEPROM}{Electrically Erasable Programmable Read-Only Memory}
\acrodef{EER}{Energy Efficiency Rating}
\acrodef{FFT}{Fast Fourier Transformation}
\acrodef{FPA}{floating Point Adder}
\acrodef{FPU}{Floating Point Unit}
\acrodef{FPM}{Floating Point Multiplier}
\acrodef{FMA}{fused multiplyÐadd}
\acrodef{FSM}{Finite State Machine}
\acrodef{GPU}{Graphics Processing Unit}
\acrodef{GPS}{Global Positioning System}
\acrodef{GSM}{Global System for Mobile Communications}
\acrodef{HC}{Hardware Counter}
\acrodef{HDL}{Hardware Description Language}
\acrodef{HPC}{High Performance Computing}
\acrodef{IC}{Integrated Circuit}
\acrodef{ILP}{integer linear programming}
\acrodef{I2C}{Inter-Integrated Circuit}
\acrodef{IQR}{interquartile range}
\acrodef{ITRS}{In\-ter\-na\-tion\-al Tech\-nolo\-gy Road\-map for Semi\-con\-duc\-tors}
\acrodef{ILP}{Instruction Level Parallelism}
\acrodef{I/O}{input/output}
\acrodef{ISA}{Instruction Set Architecture}
\acrodef{FIR}{finite impulse response}
\acrodef{JIT}{Just-In-Time}
\acrodef{JNI}{Java Native Interface}
\acrodef{LPDDR}{low power DRAM}
\acrodef{LCD}{liquid crystal display}
\acrodef{MAD}{median absolute deviation}
\acrodef{MIPJ}{millions-of-instructions-per-joule}
\acrodef{MOSFET}{metal-oxide semiconductor field-effect transistor}
\acrodef{MTTF}{Mean Time To Failure}
\acrodef{MIPS}{Microprocessor without Interlocked Pipeline Stages}
\acrodef{NIC}{Network Interface Card}
\acrodef{NIST}{National Institute of Standards and Technology}
\acrodef{NDK}{Native Development Kit}
\acrodef{NTP}{Normal Temperature and Pressure}
\acrodef{OS}{Operating System}
\acrodef{PoP}{Package-on-Package}
\acrodef{OOE}{out-of-order execution}
\acrodef{PCB}{printed circuit board}
\acrodef{PID}{proportional-integral-derivative}
\acrodef{RAM}{Random Access Memory}
\acrodef{RISC}{Reduced Instruction Set Computing}
\acrodef{ROHC}{Robust Header Compression}
\acrodef{RMS}{Root Mean Square}
\acrodef{rpm}{revolutions per minute}
\acrodef{RTL}{Register Transfer Language}
\acrodef{SIMD}{Single Instruction Multiple Data}
\acrodef{SMD}{surface mount device}
\acrodef{SoC}{Systems-on-Chip}
\acrodef{SGLP}{Super-graph Level Parallelism}
\acrodef{SLP}{Super-word Level Parallelism}
\acrodef{SPM}{Scratch-Pad Memory}
\acrodef{SVM}{State Vector Machine}
\acrodef{SRAM}{Static Random-access Memory}
\acrodef{SDRAM}{synchronous dynamic random access memory}
\acrodef{STP}{standard temperature and pressure}
\acrodef{TCP}{Transport Control Protocol}
\acrodef{TCT}{Task Completion Time}
\acrodef{TLB}{Translation Look-aside Buffer}
\acrodef{TLP}{Thread Level Parallelism}
\acrodef{TP}{Travaux Pratiques}
\acrodef{TMU}{Thermal Management Unit}
\acrodef{TTA}{Transport-Triggered Architecture}
\acrodef{UMTS}{Universal Mobile Telecommunications System}
\acrodef{VC}{Virtual Channel}
\acrodef{VM}{Virtual Machine}
\acrodef{VLSI}{Very-Large-Scale Integration}
\acrodef{VHDL}{VHSIC Hardware Description Language}
\acrodef{VLIW}{Very Long Instruction Word}
\acrodef{VM}{Virtual Machine}
\acrodef{WFL}{Weber-Fechner law}
\acrodef{WiFi}{Wireless-Fidelity}
\acrodef{WLAN}{Wireless Local Area Network}
\acrodef{WSN}{Wireless Sensor Network}

\begin{document}
\title{Modeling the Temperature Bias of Power Consumption for Nanometer-Scale CPUs in Application Processors}


\author{\IEEEauthorblockN{Karel De\,Vogeleer\IEEEauthorrefmark{1},
Gerard Memmi\IEEEauthorrefmark{1},
Pierre Jouvelot\IEEEauthorrefmark{2} and
Fabien Coelho\IEEEauthorrefmark{2}}
\IEEEauthorblockA{\IEEEauthorrefmark{1}TELECOM ParisTech -- INFRES -- CNRS LTCI - UMR 5141 -- Paris, France\\
Email: \{karel.devogeleer,gerard.memmi\}@telecom-paristech.fr}
\IEEEauthorblockA{\IEEEauthorrefmark{2}MINES ParisTech, France\\
Email: \{pierre.jouvelot,fabien.coelho\}@mines-paristech.fr}}

\maketitle

\begin{abstract}
We introduce and experimentally validate a new macro-level model of the {\small CPU} temperature/power relationship  within nanometer-scale application processors or system-on-chips.
By adopting a holistic view, this model is able to take into account many of the physical effects that occur within such systems.
Together with two algorithms described in the paper, our results can be used, for instance by engineers designing power or thermal management units, to cancel the temperature-induced bias on power measurements.
This will help them gather temperature-neutral power data while running multiple instance of their benchmarks.
Also power requirements and system failure rates can be decreased by controlling the {\small CPU}'s thermal behavior.

Even though it is usually assumed that the temperature/power relationship is exponentially related, there is however a lack of publicly available physical temperature/power measurements to back up this assumption, something our paper corrects.
Via measurements on two pertinent platforms sporting nanometer-scale application processors, we show that the power/temperature relationship is indeed very likely exponential over a 20$^\circ$C to 85$^\circ$C temperature range.
Our data suggest that, for application processors operating between 20$^\circ$C and 50$^\circ$C, a quadratic model is still accurate and a linear approximation is acceptable.
\end{abstract}
%

\section{Introduction}

Since the dawn of \acp{IC}, it is known that the currents flowing through the \acp{IC} produce heat dissipation proportional to $I^2R$, where $R$ is the resistance and $I$ the electric current.
The first law of thermodynamics states that in steady operation the energy input of a system is equal to the energy output of the system.
Thus, in the absence of other energy interactions, most energy leaves an \ac{IC} in the form of heat, resulting from currents flowing through electrical elements~\cite{cengel2010heat}.
Therefore the \ac{IC}'s heat dissipation is proportional to its power consumption.
Moreover the \ac{IC} will exhibit transient thermal behavior, where the time-frame depends on its heat capacity.
The transient thermal behavior is more lasting for systems with larger heat capacities than systems with smaller heat capacities.
Such systems can be thought of as {\small RC}-circuits, i.e., a low-pass filter, where the temperature of the system is proportional to the voltage over the capacitors~\cite{10.1109/L-CA.2003.5}.
In reality the (transient) thermal behavior is more complicated, because the resistance $R$ and the current $I$ are known to depend on the \ac{IC}'s temperature $T$, which is non-uniform over the \ac{IC} and time-dependent.
Generally speaking, it is shown that the power dissipation grows super-linear with the temperature of the system.

The \ac{IC}'s heat generation process is sometimes also referred to as \emph{internal heat conversion}.
It is the \emph{temperature/power relationship}, i.e., the relationship between the internal heat conversion and the temperature, that is the subject of this work.
In particular we study the thermal behavior of three application processors for embedded systems such as smartphones, netbooks, or tablets.

Elevated temperatures have adverse effects on \acp{IC}.
The reliability of electronic products can be influenced by spatial or temporal gradients, or absolute temperatures~\cite{lall1997influence}.
The \ac{MTTF} of electronic equipment increases exponentially temperature: possible causes of failure are electron migration, chemical reactions, dielectric breakdown, or creep in the bonding materials~\cite{cengel2010heat}.
For safety reasons, the system's temperature is also limited.
Smartphones are often thermally capped around 50$^\circ$C so that its users don't burn any body parts, but also to maximize battery life and up-time.
Such applications employ \acp{TMU}, which are able to throttle or scale the system so that stringent thermal constraints are met.
Complex \acp{CPU} may employ hard-wired \acp{TMU}, e.g., some Intel chip sets~\cite{1564363,4405718}, whereas \ac{TMU} software implementations are frequently seen in embedded devices.
For \acp{TMU}, it is important to understand the transient thermal behavior of systems.
The transient thermal behavior of actively cooled systems can be well described via an exponential relationship.
But, for passively cooled systems such as smartphones, wireless sensors, appliances or vehicles, the exponential assumption does not hold.
Passively cooled systems rely on cooling via conduction, natural convection, radiation and the \ac{TMU}, which exhibit non-linear properties.
To be optimally effective, \acp{TMU} must therefore have an adequate understanding of the system.
This includes knowledge about its transient thermal behavior, the internal heat conversion and its temperature dependency.

There are several factors that affect the internal heat conversion dependency on temperature.
Some manifest at the micro-level; others result from macro-level effects.
The most known contributor to the temperature dependence of internal heat conversion are the \emph{transistor leakage currents}.
Leakage currents effects are inherent to silicon-based \acp{MOSFET} with which \acp{CPU} are built.
It is well known that these currents are temperature-dependent~\cite{Liu:M00/48}.
On a macro level, a \acs{CPU} is fed by a voltage regulator which also contains a cohort of transistors and other electrical elements.
Given the temperature-dependent behavior of transistors, the voltage regulator will be temperature-dependent as well.
Moreover, the physical properties such as the \emph{electrical resistance} and \emph{thermal resistance} of the materials that compose the \acs{CPU} are themselves temperature-dependent.
It is the combined effect of such phenomena that result in the non-linear thermal behavior of systems.
Many theoretical studies focus solely on the effects of leakage currents, a priori neglecting any other sources responsible for temperature-dependent behavior.

From a measurement perspective, it is also important to understand the implications of the internal heat conversion temperature dependency.
The reproducibility of accurate power measurements are challenging by virtue of the transient thermal behavior of \acp{IC}.
More specifically, for a fixed benchmark, a \acs{CPU} power measurement will yield different values at different \acs{CPU} temperatures.
For the sake of accuracy and fair comparison between different power measurements, it is thus of vital importance to control or cancel the effects of transient and static thermal behaviors.

The main contributions of this paper are:
\begin{itemize}
  \item experimental evidence that the aggregated behavior of the temperature/power relationship is very likely exponential, while, for small temperature variations, quadratic and linear approximations may be adequate;
  \item a new model that estimates the influence of temperature on an application processor's \ac{CPU} power consumption, for a given frequency and active core count;
  \item a simple method to remove the temperature bias from a power measurement trace.
\end{itemize}

The rest of the paper is organized as follows.
Section~\ref{sec:contributors} elaborates on the processes that may influence the temperature-dependent behavior of the internal heat conversion.
In Section~\ref{sec:literature}, examples of temperature/data traces found in the literature are listed; they illustrates the internal heat conversion.
Section~\ref{sec:testbed} describes the testbed and measurement methodology we used to gauge the internal heat conversion in our own measurements.
Section~\ref{sec:measurements} presents the temperature-power traces we collected and their analyses, plus a generalized power model.
Section~\ref{sec:real-life} presents a real-life application of the results from the previous Section.
Section~\ref{sec:futurework} sheds some light upon possible future work, mainly on how to improve the accuracy of the measurements.
We conclude in Section~\ref{sec:conclusion}.




\section{Contributors to Temperature Fluctuations}
\label{sec:contributors}

The \emph{total power consumption} $\Pcpu$ of a \acs{CPU} is linked to all the currents that flow within the \acs{CPU} accounted for by different physical processes~\cite{paper:karel:warsaw}.
Firstly, the \emph{dynamic power} $\Pdyn$ is the power used to charge and discharge the capacities that drive the logic gates within the \acs{CPU}.
Each time the \acs{CPU} changes its state, i.e., gates are toggled, energy is required or flushed to maintain this new state.
Moreover, during the state changes of the logic gates, the transistors inside may conduct simultaneously for a very brief moment, shorting the supply to ground.
Currents flowing as part of this process contribute to the \emph{short-circuit power} consumption $\Psh$.
Also, leakage currents flow though the transistors as a result of their non-ideal behavior resulting in a leakage power $\Pleak$.
The total \acs{CPU} power consumption can then be thought of as the sum of the mentioned processes, whence
\begin{equation}
\Pcpu = \Pdyn + \Psh + \Pleak.
\end{equation}
$\Pleak$ is known to be temperature-dependent and well defined for an isolated transistor.
$\Pdyn$ and $\Psh$ are related to changing physical material properties resulting from temperature fluctuations.
Furthermore, $\Pcpu$ is itself temperature-sensitive as the voltage supply level may be temperature-dependent.
The total internal heat conversion is the combined effect of the mentioned processes whose relations are not totally clear, and moreover could lag in time.
Therefore, further on, we approach the modeling of the temperature/power relationship at a macro level such that the aggregated behavior of the mentioned, and tentatively other, physical processes are captured.

\subsection{Leakage Currents}

Among the multiple sources of leakage in \ac{MOSFET} transistors, the sub-threshold leakage current, gate leakage and \ac{BTBT} dominate the others for sub-100\,nm technologies~\cite{1468683,Xu:2004:LCE:1044241.1044258}.
Leakage current models, e.g., as incorporated in the \ac{BSIM}~\cite{Liu:M00/48}, are accurate, nevertheless complex, since they depend on multiple parameters.
Detailed knowledge of the transistors are necessary to assess the precise magnitude of the leakage currents, e.g., dimensions, materials and terminal voltages.
Even more, when transistors are stacked, e.g., in logic-gates, leakage currents may be amplified throughout the stack~\cite{Xu:2004:LCE:1044241.1044258}. 
As a \acs{CPU} changes state each clock cycle, keeping track of the exact leakage current over time may practically be a daunting task.

As the physical dimensions of transistors shrink each generation and materials used are optimized, one should be careful to assess the magnitude of leakage currents based on past research.
It has been shown that, for example, gate leakage becomes more prominent when transistor dimensions shrink~\cite{1683982}. 
Mostly sub-threshold leakage currents are accounted for in previous research, which was not necessarily a wrong assumption for the larger transistor sizes studied in the past.
However, as a result of composed leakage current effects, we are poised to develop a model that captures this aggregated behavior over time and its temperature dependency, on a macro level for all transistors in the \acs{CPU}.


Often course-grained models are inspired by the intrinsic behavior of a single transistor's leakage current.
Table~\ref{table:leakagecurrent} provides an overview of models found in the literature.
\begin{table}[!t]
  \addtolength{\abovecaptionskip}{1.25em}
  \centering
  \caption{Literature models capturing the coarse-grained behavior of leakage currents.
	    $a_*$ are scalars, $T$ is the {\smaller CPU} temperature.\label{table:leakagecurrent}}
  \begin{tabular}{|l|l|}
    \hline
    {\sc Authors}		& {\sc Model}\\\hline
    Su~\cite{Su:2003:FCL:871506.871529}		& $\Pl \propto a_2 T^2 + a_1 T + a_0$ \\
    Liao~\cite{Liao:2006:TSV:2298535.2301211} 	& $\Pl \propto a_1 T^2 e^{a_0/T}$ \\
    Liao~\cite{Liao:2002:LPM:774572.774677} 	& $\Pl \propto a_2 e^{-a_0/(T-a_1)}$ \\
    Liu~\cite{Liu:2007:ATI:1266366.1266701}	& $\Pl \propto a_2 + a_2 (T-a_1)$ \\
    Ferr\'e-Sinha~\cite{814859}~\cite{Sinha:2001:JWB:378239.378467} & $\Pl \propto a_1 e^{a_0/T}$			\\
    Skadron~\cite{Skadron:2004:TMM:980152.980157}	& $\Pl \propto a_1 T^2 e^{-a_0/T}$ \\
    Skadron~\cite{Zhang03hotleakage:a}		& $\Pl \propto a_2 (1-e^{a_1/T})e^{a_0/T}$ \\\hline
  \end{tabular}
\end{table}
Liao et al.~\cite{Liao:2006:TSV:2298535.2301211} stated that for their 65\,nm benchmark the sub-threshold and gate leakages dominate the leakage process.
Only the former is temperature-dependent, the authors claim.
In another paper by Liao et al.~\cite{Liao:2002:LPM:774572.774677} a power consumption model for adders was presented.
The temperature-dependent part of the power model looks slightly different from their previous work. 
Skadron et al.~\cite{Skadron:2004:TMM:980152.980157} deducted a relationship between the leakage power $\Pleak$ and dynamic power $\Pdyn$ based on \acf{ITRS} power traces.
It can be observed that their equation is based on the sub-threshold leakage current.
In another publication, Skadron et al.~\cite{Zhang03hotleakage:a} adopted the exact formulation of the sub-threshold leakage currents for the transistor in the \emph{off} state.
Liu et al.~\cite{Liu:2007:ATI:1266366.1266701} put forward a linearized leakage current equation based again on the gate and threshold leakage current, which was studied via {\small SPICE} simulations.
Yet, in their humble attempt to model the thermal behavior of leakage currents with finite elements, they forgot to account for the inflating power consumption due to leakage currents.
Su et al.~\cite{Su:2003:FCL:871506.871529} modeled the leakage currents of so called \emph{standard cells} using {\small SPICE} and custom thermal simulations.
The authors identified a satisfactory quadratic correlation between temperature and leakage currents.
Ferr\'e and Figueras~\cite{814859} and also Sinha and Chandrakasan~\cite{Sinha:2001:JWB:378239.378467}, based on the sub-threshold leakage current, assumed a pure exponential relationship between temperature and leakage.

Most authors assert that they were able to model the leakage currents adequately on their dedicated testbed.
This suggests that leakage currents may very well be application-specific, i.e., for given transistor dimensions and materials etc.
Measuring the leakage current on a real testbed is not a straightforward task; therefore simulations are often employed to quantify its magnitude.
Yet, we can get a glimpse of its behavior by adjusting temperature levels.


\subsection{Voltage Regulators}
Each \acs{CPU} has a voltage regulator that supplies the \acs{CPU} with a constant voltage supply.
The voltage regulators of \ac{DVFS}-enabled \acp{CPU} can alter the magnitude of the supplied voltage on demand, though with a small transition delay.
A voltage regulator is built up from capacitors, inductors and transistors.
Therefore the voltage regulator will also supply a voltage where its magnitude depends on the temperature.
The resulting macro-level effect is that, for a fixed resistance, the current supplied to a resistor will increase and augment the internal heat conversion.
For example, via Ohm's law, we calculate that, if a 1.5 voltage drop over a 1\,k$\Omega$ resistor increases by 1\%, the resistor's power dissipation will increase by 2\%.

The temperature sensitivity of the voltage regulator is usually listed in its datasheet.
The {\small S2MPS11} voltage regulator of our {\small ODROID} testbed (defined in the sequel) is unfortunately not at our disposal.
But we can estimate its temperature drift via the onboard {\small INA231} voltage sensor.
We observed in the most extreme case\footnote{All {\smaller CPU} cores active at maximum frequency.} a voltage rise of 1.25\,V to 1.265\,V between 30\dgr~and 90\dgr~for the A15 processor, which corresponds to a 1.2\% voltage increase.
The maximum gain error for the {\small INA231} is listed to be 0.5\%.
This leaves us with an estimated voltage regulator temperature drift of around 0.8\%.
For the more energy-efficient A7 processor, in idle mode a 0.25\% rise was noted in the most extreme case.
The quantization noise and gain error, however, render the latter observation unreliable.
Regardless of the large measurement errors, we have an indication that temperature may affect the voltage regulator's output.

In general, the voltage regulator may not always be exposed to the full temperature swings stemming from switching logic inside the \acs{CPU}.
This depends on the relative distance of the voltage regulator to the \acs{CPU}.
As a result, during peak \acs{CPU} activity, escalating internal heat conversion, resulting from leakage current swells, will kick in faster than the increased dissipation from the inflating voltage supply.
This is a result of the finite propagation time of heat between the \acs{CPU}'s logic and the voltage regulator.
As an illustration, in both our testbeds the voltage regulator ({\small S2MPS11} and {\small MAX8997}) is located about 1\,cm away from the actual application processor.
Practically this implies that the transient thermal behaviors of a \acs{CPU} when the whole system is heated homogeneously, e.g., while being exposed to the sun, and when the temperature rise in the \acs{CPU} emanates from the execution of a job will look different.

\subsection{Physical Properties of the CPU}

Physical properties of the materials that constitute the \acs{CPU} and \ac{PCB} are temperature-dependent, including the electrical resistance and thermal diffusivity.
For example the resistivity of copper and aluminum increases about 12\% between 0\dgr~and 50\dgr.
As an other practical illustration, note that a thick film resistor's, e.g. \ac{SMD}, electrical resistivity decreases 1\% over 50\dgr.
The electrical resistivity of semiconductors typically decreases with rising temperatures.
Similarly, the thermal conductivity of both copper and aluminum changes about 0.9\% between 0\dgr~and 50\dgr.

Even though these physical properties are temperature-dependent, they probably have a small influence on the temperature dependence of the internal heat conversion.

\section{Temperature/Power Models in the Literature}
\label{sec:literature}

With the objective of reaching adequate performance within temperature constraints, \ac{DVFS} controllers may employ temperature/power models.
Also for \acp{TMU} it may be useful to understand the thermal behavior of the temperature/power relationship.
Computer energy consumption decompositions account often for the leakage currents where the temperature/power dependency is referenced.
A summary of temperature/power relationship models found in the literature is listed below.

Weissel and Bellosa~\cite{weissel04thermalmanagement} developed a \ac{TMU} for data center computers.
Based on a handful of temperature/power measurements in a limited temperature range (35\dgr~to 60\dgr), they assumed the temperature/power relationship to be quadratic, quasi linear.
The accuracy of the fitting is however questionable.
Hanumaiah and Vrudhula~\cite{10.1109/TC.2011.156} developed a \ac{DVFS} controller for systems with hard real-time and temperature constraints.
They employ a linearized version of the exponential temperature/power assumption, which was based on the \BSIM leakage current models.
The authors also assume that the power increases linearly with the supply voltage.
The temperature in their experiments ranged between 35\dgr~and 110\dgr.
While studying the thermal response to \ac{DVFS} of an Intel Pentium M processor, Hansom et al.~\cite{5514321} assumed a linear relationship between power and temperature.
The temperature ranged between 20\dgr~and 55\dgr~in their experiments.
Sinha and Chandrakasan~\cite{Sinha:2001:JWB:378239.378467} decomposed the energy consumption of a StrongARM platform.
Based on the \BSIM definition of the sub-threshold leakage current, they proposed an exponential relationship to represent the leakage current. 
The temperature is implicitly referenced in the denominator of the natural exponent; other than that the temperature is not mentioned.
Liao et al.~\cite{Liao:2006:TSV:2298535.2301211}, in their simulations, assess a \acs{CPU}'s performance.
The authors also assume an exponential behavior based on the sub-threshold leakage current. 
The temperature in their experiments ranges between 65\dgr~and 110\dgr.
Singh et al.~\cite{Singh:2009:RTP:1577129.1577137} attempted to model an {\small AMD} \acs{CPU}'s power consumption based on a subset of performance counters.
Temperature/power traces are shown but only with relative figures.
Their traces show that during the execution of some benchmarks the \acs{CPU}'s temperature inflates about 20\%, resulting in a 10\% power increase.
With a bit of good-will, a super linear relationship can be identified.
Ikebuchi et al.~\cite{DBLP:conf/aspdac/IkebuchiSKKZASKHUMUTNNK10} show temperature/power traces for their Geyser-1 {\small MIPS} \acs{CPU}.
The temperature/power relationship, measured between 20\dgr~and 80\dgr, shows a clear exponential relationship.
The authors also show that with the help of power-gating the effects of leakage currents on power consumption can be diminished.

The works listed above show temperature/power traces for at most three benchmarks and for specific \acs{CPU} settings, mostly for illustrative purposes.
Usually high-performance {\small MIPS} processors are targeted as the objects of study, as part of large server farms.
Based on elaborate measurements described further on, we identify an experimental temperature/power relationship for different \acs{CPU} configurations and loads for our application processors.
Such application processors are expected to function in embedded systems, e.g., smartphones, appliances, vehicles or smart sensors.


\section{Testbed}
\label{sec:testbed}

We used the following two platforms to collect temperature/power traces.
The first is a Samsung Galaxy S2 sporting the Samsung Exynos 4 \ac{SoC} 45\,nm dual-core, and the second, a Hardkernel {\small ODROID XU+E} featuring the Samsung Exynos 5 \ac{SoC} 28\,nm quad-core. 
The Galaxy includes an A9 Cortex processor, whereas the {\small ODROID} has both an A7 and an A15 Cortex processor on the same die.
The two platforms were running a custom compiled Linux kernel.
The frequency scaling governor was set to operate in \emph{userspace} mode to prevent frequency and voltage scaling on-the-fly.

The {\small ODROID} has onboard power sensors with an accuracy around 1.25\,mW.
To measure the power consumption of the Galaxy we replaced its battery with a power supply (Monsoon Power Monitor) that samples the power with about 1\,mW accuracy.
The temperature on both platforms was measured via onboard temperature sensors with a 1$^\circ$C accuracy.
Power and temperature samples were collected at a rate of 5\,Hz.
We applied forced cooling and forced heating to the \acp{SoC} packaging (including the \acs{CPU}) to force its temperature up and down.



During the trace recording a constant load is applied to one or more cores of the \acp{CPU}.
The Galaxy was loaded with 4096\,kB bit-reverse calculations.
We used the Gold-Rader implementation of the bit-reverse algorithm, part of the ubiquitous \ac{FFT} algorithm, which rearranges deterministically elements in an array.
The {\small ODROID} spun over the square root function from the default math library.
The root calculations were forked up to four times to assess the temperature/power impact of the four cores in the A7 and A15 processors; the inactive cores were not hot-plugged.
On the A9 platform we only enabled one of the two cores; the other core was unplugged.
Some other \ac{CPU} peripherals were disabled, including the screen and camera, to minimize noise in the power measurements.
It must be noted that the benchmarks ran on top of an \ac{OS}, so there must be some power accounted to the system's overhead.

\section{Temperature/Power Modeling}
\label{sec:measurements}

Because of the temperature dependency of some currents flowing though the \acs{CPU}, the \acs{CPU} power consumption will inflate for increasing silicon temperatures. We describe our experiments and the temperature/power model we deduced from them.

\subsection{Experiments}

\begin{figure}[t!]
  \begin{center}
    \input{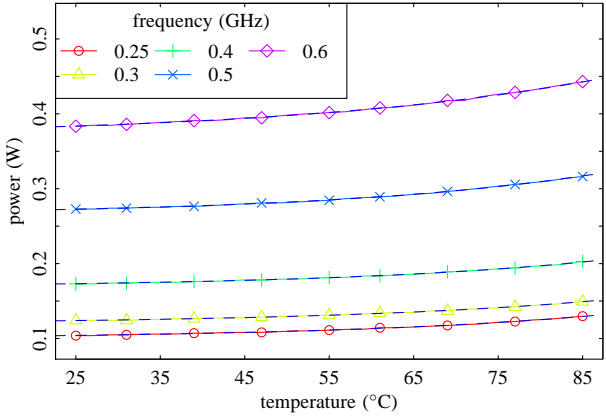}
  \end{center}
  \caption{Temperature/power traces for the A7 processor with three active cores at different frequencies.
	    The dashed blue lines are the fitted exponential curves as described in Equation~\ref{eq:exponent}.}
  \label{fig:overview-a7}
\end{figure}

\begin{figure}[t!]
  \begin{center}
    \input{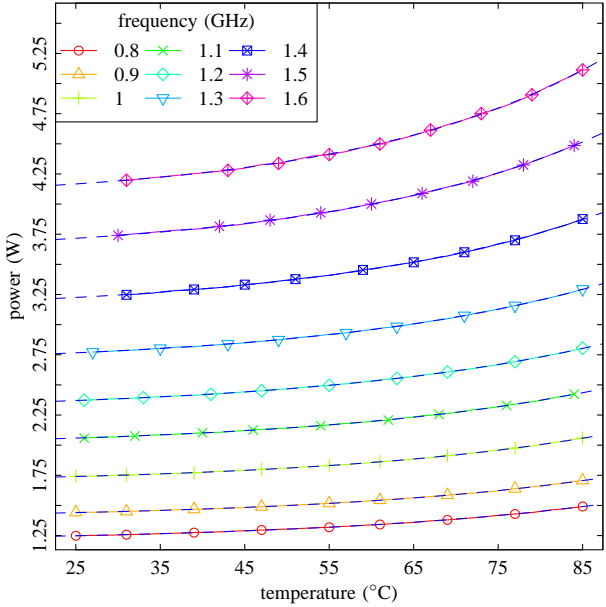}
  \end{center}
  \caption{Temperature/power traces for the A15 processor with four active cores at different frequencies.
	    The dashed blue lines are the fitted exponential curves as described in Equation~\ref{eq:exponent}.}
  \label{fig:overview-a15}
\end{figure}

\begin{figure*}[!t]
  \addtolength{\abovecaptionskip}{1.5em}
  \centerline{\subfloat[A7 processor - 500\,MHz - \#pr=3]{\input{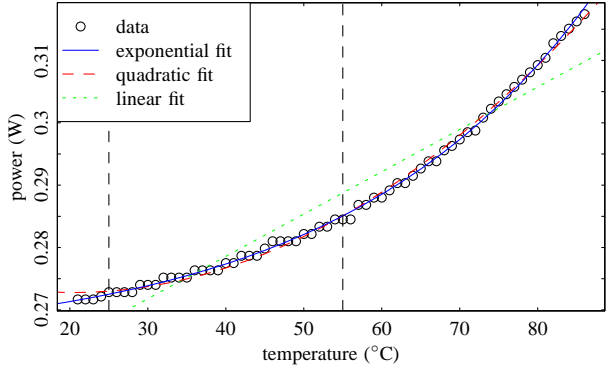}%
    \label{fig:exponential-a7}}
    \hfil
    \subfloat[A15 processor - 1.2\,GHz - \#pr=4]{\input{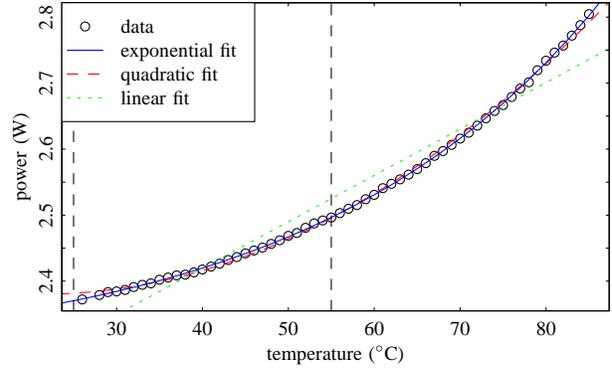}%
    \label{fig:exponential-15}}}
   \caption{Temperature/power relationship as measured on the A7 and A15 processors for three and four active cores (\#pr), respectively.
	      The traces for the other processors and configurations show similar behavior.
	      Note the quantization noise in the case of the A7 processor, most prevalent at low temperatures.}
   \label{fig_sim}
\end{figure*}

We artificially swept the temperature between 25$^\circ$C and 85$^\circ$C for the A7 and A15 processors; for the A9 the temperature was swept between 25$^\circ$C and 55$^\circ$C.
We measured the power consumption and temperature of the A7 between 250\,MHz and 600\,MHz, the A9 between 200\,MHz and 1.6\,GHz, and the A15 between 0.8\,GHz and 1.6\,GHz.
Excerpts of the traces are shown in Figure~\ref{fig:overview-a7} and Figure~\ref{fig:overview-a15}.
We fitted all the traces with three types of curves to assess their applicability:
\begin{enumerate}
 \item linear curve:
    \begin{equation}
      P = a_{1} T + a_{0};
    \end{equation}
 \item quadratic curve:
    \begin{equation}
      P = a_{2}T^2 + a_{1} T + a_{0};
    \end{equation}
 \item exponential curve:
    \begin{equation}
      P = e^{(T-a_{1})/a_{2}} + a_{0}.\label{eq:exponent}
    \end{equation}
\end{enumerate}
where $a_*$ are scalars, to be defined via fitting.
Figures~\ref{fig:exponential-a7} and~\ref{fig:exponential-15} show single temperature/power traces for the A7 and A15 processors, respectively.
Almost all traces look similar to these two examples; therefore, and for space reasons, we do not show all curves and present aggregated fitting errors.
Aggregated fitting errors are given in Table~\ref{table:fiterrorswhole} for fitting over the 25$^\circ$C to 85$^\circ$C temperature range.
\begin{table}[t!]
\addtolength{\abovecaptionskip}{-1em}
\caption{Aggregated temperature/power linear, quadratic ({\sc quad}) and exponential ({\sc expo}) fitting errors over the temperature range 25$^\circ$C to 85$^\circ$C for the A7 and A15 processors ({\sc pr}) and a given active core count ({\sc \# co}).}
\begin{center}
\begin{tabular}{|c|c|c|c|c|}\hline
 {\sc pr} & {\sc \# co} & {\sc linear} & {\sc quad} & {\sc expo} \\\hline\hline
A7  &  1  &  0.190058  &  0.048509  &  0.039109 \\
A7  &  2  &  0.126700  &  0.030710  &  0.023524 \\
A7  &  3  &  0.111039  &  0.026028  &  0.016972 \\
A7  &  4  &  0.103021  &  0.023661  &  0.014973 \\\hline
A15 &  1  &  0.120828  &  0.021422  &  0.008501 \\
A15 &  2  &  0.098198  &  0.017627  &  0.007960 \\
A15 &  3  &  0.085662  &  0.014614  &  0.006146 \\
A15 &  4  &  0.084010  &  0.014459  &  0.005787 \\\hline
\end{tabular}
\end{center}
\label{table:fiterrorswhole}
\end{table}%
The fitting errors are aggregated over all the traces measured with the same active core count.
The fitting errors were computed as follows:
\begin{equation}
	\mathrm{error} = \sqrt{\sum_i \left(\frac{\tilde{y}_i - y_i}{y_i}\right)^2},
\end{equation}
where $y$ are the measured values and $\tilde{y}$, the model.

We observe that the sum of errors for the quadratic case are on the average 2.5 times larger than the exponential fit errors, and the linear fit errors are about six times larger than the quadratic.
The $p$-values of the \emph{sign test} between the three models on the A7 and A15 stay well below the 0.01 significance level, confirming that the exponential is a significantly better fit than the quadratic, while the latter is significantly better than the linear fit.
Thus, the exponential fit would be the most representative of the three proposals.
Figures~\ref{fig:exponential-a7} and~\ref{fig:exponential-15} back up this observation.
Indeed, the exponential fit seems to follow the measurements very well.
The quadratic curve overestimates the power for the lower temperatures but performs very well for larger temperatures.
The linear curve does not adequately represent the temperature/power relationship in this temperature range compared to the other two proposals.


Let's see how the curves behave in the pertinent temperature range between 25$^\circ$C and 55$^\circ$C.
Aggregated errors are given in Table~\ref{table:fiterrorshalf}.
\begin{table}[t!]
\caption{Aggregated temperature/power linear, quadratic (quad) and exponential (expo) fitting errors ($10^{-3}$) over the temperature range 25$^\circ$C to 55$^\circ$C for the A7 and A15 processors (pr) and a given active core count (\# co).}
\begin{center}
\begin{tabular}{|c|c|c|c|c|}\hline
{\sc pr} & {\sc \# co} & {\sc linear} & {\sc quad} & {\sc expo} \\\hline\hline
A9  &  1  &  0.197413  &  0.147489  &  0.158332 \\\hline
A15 &  1  &  9.519750  &  4.648533  &  4.590021 \\
A15 &  2  &  8.536068  &  4.558667  &  4.596187 \\
A15 &  3  &  6.327271  &  3.244339  &  3.206688 \\
A15 &  4  &  6.318451  &  3.000202  &  2.96603  \\\hline
\end{tabular}
\end{center}
\label{table:fiterrorshalf}
\end{table}%
Due to the quantization noise within this temperature span, the fitting for the A7 doesn't always converge properly.
This hampers the fitting process and renders the results unreliable; therefore we omit its analysis here.
For the A15 processor, we observe that the competitive advantage of the exponential curve has shrunk.
The \emph{sign test} significantly favors the quadratic curve over the linear curve.
The quadric curve in the 25$^\circ$C--55$^\circ$C temperature range is, however, as good as the exponential curve based on the same sign test.
In the case of the A9 processor, based on the aggregated errors, the quadratic curve presents itself as the best fit, but it is as good as the exponential according to the sign test with a 0.01 significance level.
Nevertheless, the linear curve is also a good match compared to the quadratic and the exponential.
We must note, however, that the A9 traces suffer from the so-called \emph{distant sensor syndrome}, described in Section~\ref{sec:futurework}.

Even tough the curves are most likely exponential, in this limited temperature range the quadratic curve performs as well as the exponential curve.
The performance of the linear curve is also acceptable.
This is a positive conclusion for \acp{TMU} and previous research that assumed a linear or quadratic relationship between temperature and power.
Analytical derivations can be notably simplified by virtue of said assumptions.


\subsection{General Temperature/Power Model and its Error Analysis}

For diverse purposes such as simulations, among others, it is useful to define the scalars $a_*$ in Equation~\ref{eq:exponent} for arbitrary frequencies and active core count.
From analyzing the A7 and A15 traces it appears that $a_2$ is seemingly constant for all measurement data.
Observing the fitted values for $a_1$ reveals that these are linearly correlated with frequency and active core count.
For the $a_0$ case, the values are quadratically correlated with the frequency and linearly with the active core count.
Moreover, the lines linking the $a_2$ values for a fixed frequency extrapolated seem to converge towards a single point on the abscissa.
Thus we suggest to use the following expressions for $a_0$ and $a_1$:
\begin{eqnarray}
g_s & = &  m_1 + m_2 f + m_3 f^2 \nonumber\\
g_o & = &  g_s/m_4 \nonumber\\
a_0 & = &  g_s c + g_o \\
a_1 & = &  m_5 f + m_6 + (5-c) m_7,
\end{eqnarray}
where $f$ is the \acs{CPU} frequency, $c$, the active core count and $m_*$, case-specific scalars.
For both processors we observed that $m_4$ and $m_7$ are approximately equal.
We have also tried to see whether the \acs{CPU} supply voltage may be correlated with the $a_*$ scalars; however, we haven't identified a satisfactory correlation.

A prototype implementation of the power models for the A7 and A15 is given in Figure~\ref{algo:a7} and Figure~\ref{algo:a15}, respectively, which can be copy-pasted directly into any simulation software.
\begin{figure}[t!]
  \addtolength{\abovecaptionskip}{1em}
  \begin{algorithmic} 
    \Procedure{Power.A7}{$T,f,c$}  
    \State $g_s \gets 0.028 - 0.093 f + 0.371 f^2$  
    \State $g_o \gets g_s / 2.202$
    \State $a_0 \gets g_s c + g_o$
    \State $a_1 \gets -38.242 f + 187.668 + (5-c) 8.430$
    \State $a_2  \gets 33.105$
    \State \textbf{return} $\exp((T-a_1)/a_2)+a_0$ 
    \EndProcedure  
  \end{algorithmic}  
  \caption{Algorithm to compute an A7 power estimate given the temperature $T$ ($^\circ$C), CPU frequency $f$ (GHz) and active core count $c$ (1:4).}
  \label{algo:a7}
\end{figure}
\begin{figure}[t!]
  \addtolength{\abovecaptionskip}{1em}
  \begin{algorithmic}
    \Procedure{Power.A15}{$T,f,c$}  
    \State $g_s \gets 0.220 - 0.315 f + 0.467 f^2$  
    \State $g_o \gets g_s / 2.202$
    \State $a_0 \gets g_s c + g_o$
    \State $a_1 \gets -56.652 f + 165.896 + (5-c) 8.430$
    \State $a_2  \gets 33.105$
    \State \textbf{return} $\exp((T-a_1)/a_2)+a_0$ 
    \EndProcedure  
  \end{algorithmic}  
  \caption{Algorithm to compute an A15 power estimate given the temperature $T$ ($^\circ$C), CPU frequency $f$ (GHz), and active core count $c$ (1:4).}
  \label{algo:a15}  
\end{figure} 
Based on the collected traces, the A15 power model above shows a median error of 1.19\% and a maximum error of 7.11\%; for the A7 processor the median error is 2.89\% and the maximum is 8.31\%.
In absolute terms, both models deviate about equally from the measurements, but as the A7 consumes less power its relative error is larger.
The presented errors are not negligible.
After analyzing our data we have identified so far three sources that introduce errors/noise: the initial temperature conditions that vary for each trace, the temperature sensor noise and the non-uniform temperature gradient while heating.

\section{Temperature-Bias Cancellation in Power Consumption Measurements}
\label{sec:real-life}

In the previous section we have shown that the temperature/power relationship is very likely exponential.
A linear or quadratic relationship is adequate as well for a limited temperature range.
In this section we provide one application of these results by showing how to improve power measurement accuracy by canceling the inflations stemming from temperature fluctuations.

\begin{figure*}[t!]
  \begin{center}
    \input{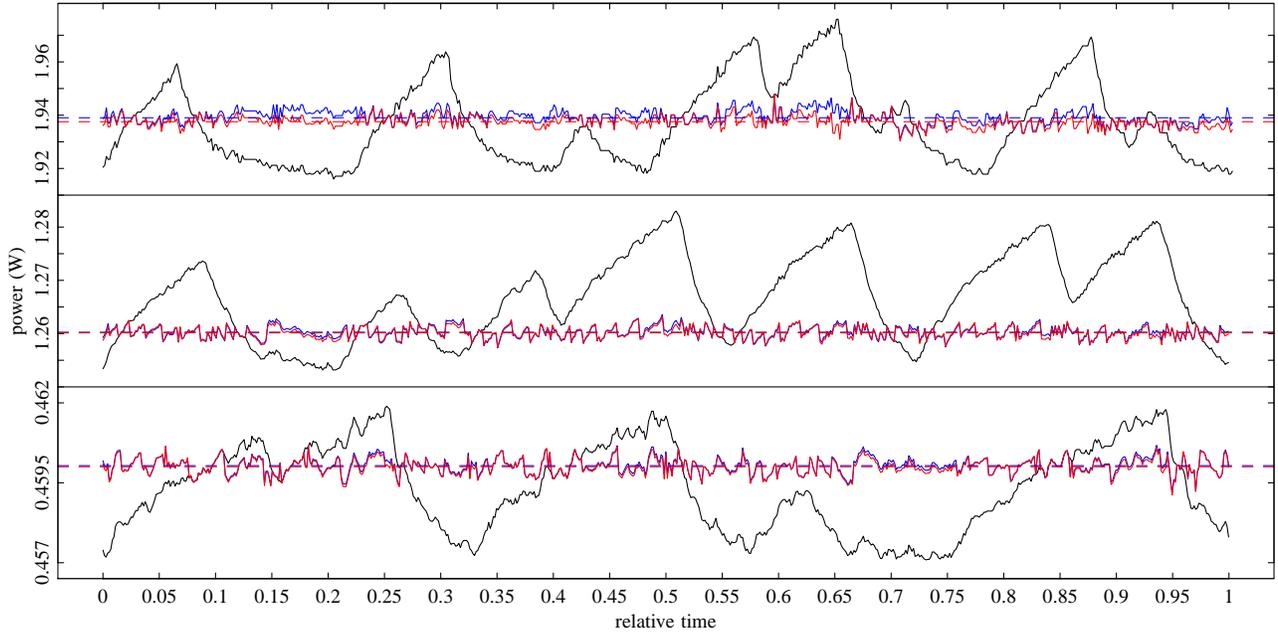}
  \end{center}
  \caption{Power/time traces (black) of (top) the A15 running at 1.3\,GHz with 3 active cores, (middle) the A15 running at 0.9\,GHz and (bottom) the A7 running at 0.6\,GHz with 4 active cores.
	    The blue and red lines are the transformed power traces in the case of linear and quadratic temperature/power approximations, respectively.
	    A random reference temperature was chosen at (from top to bottom) 40$^\circ$C, 46$^\circ$C and 55$^\circ$C.}
  \label{fig:waves}
\end{figure*}

Figure~\ref{fig:waves} shows three examples of actual power measurement traces; the temperature was also recorded.
Our goal is to convert all measured power samples as if they were measured at a fixed arbitrary \emph{reference temperature}.
Based on the temperature/power relationship fittings provided in the previous section, we may assume that there must exist a transformation function (linear/quadratic/exponential) such that the transformed power is constant\footnote{In practice, due to noisy measurements, the variance of the power measurements is to be minimized.}, i.e., normalized w.r.t. a reference temperature.
As a result the power traces as shown in Figure~\ref{fig:waves} should appear flat after the transformation.
A linear relationship between temperature and power ($P=\eta_1 T + \eta_0$) yields the following transformation function, similar to differential approximation:
\begin{eqnarray}
P_r - \eta_1 T_r & = & P_m - \eta_1 T_m  \nonumber\\
P_r  & = & P_m + \eta_1 \Delta T,
\end{eqnarray}
where $P_r$ is the power at the reference temperature, $T_r$, the reference temperature, $P_m$ and $T_m$, the measured power and temperature and $\Delta T = T_r - T_m$.
Similarly, a quadratic temperature/relationship ($P=\eta_2 T^2 + \eta_1 T + \eta_0$) yields the following power transformation function:
\begin{equation}
P_r = P_m + \eta_2 (T_r^2-T_m^2)+ \eta_1 \Delta T.
\end{equation}

To find the optimal transformation function it is not necessary to know the  \acs{CPU}'s precise thermal behavior.
A linear or quadratic regression between temperature and power of the collected traces suffices to obtain the $\eta_*$ values.
As stated before, this approach is appropriate when the testbed temperature variations are no more than 30$^\circ$C; otherwise one needs to resort to exponential fits to maintain acceptable accuracy.

Figure~\ref{fig:waves} shows random irregular power traces and their resulting linear (blue) and quadratic (red) power transformations.
As can be observed, the jerky power traces are converted into stable traces, except for the presence of some noise.
The temperature noise and inaccuracy is a known problem for \acp{TMU}~\cite{Kong:2012:RTM:2187671.2187675}.
The most important feature of our power transformation is that the arbitrary distribution is transformed into a symmetric distribution.
Table~\ref{table:fittingper} shows an overview of the transformation performance for different reference temperatures.
\begin{table}[t!]
  \caption{Performance metrics of the power transformation for different reference temperatures $T$ ($^\circ$C) and processor configurations, as stated in Figure~\ref{fig:waves}.
  The measured maximum power inflation is provided ({\sc afl}), as well as the relative fluctuation ({\sc fl}) for the linear ({\sc l}) and quadratic cases ({\sc q}).
  The ratio between the median and the mean is also provided ({\sc rat}) to assess the measurement distribution's symmetry.}
  \begin{center}
    \begin{tabular}{*6{|c}|}\hline
    \multicolumn{6}{|c|}{\sc A15 @ 1.3\,GHz}\\\hline
    {\smaller $T$} & {\sc afl} & {\sc fl-l} & {\sc fl-q} & {\sc rat-l} & {\sc rat-q} \\\hline
    35  &  1.21  &  0.297  &  0.311  &  0.753$\cdot10^{-3}$  &  0.403$\cdot10^{-3}$ \\
    38  &  1.21  &  0.295  &  0.310  &  0.750$\cdot10^{-3}$  &  0.401$\cdot10^{-3}$ \\
    41  &  1.21  &  0.294  &  0.309  &  0.747$\cdot10^{-3}$  &  0.400$\cdot10^{-3}$ \\\hline
    \multicolumn{6}{|c|}{\sc A15 @ 0.9\,GHz}\\\hline
    {\smaller $T$} & {\sc afl} & {\sc fl-l} & {\sc fl-q} & {\sc rat-l} & {\sc rat-q} \\\hline
    43  &  1.80  &  0.231  &  0.212  &  -1.706$\cdot10^{-3}$  &  -2.591$\cdot10^{-3}$ \\
    46  &  1.80  &  0.229  &  0.211  &  -1.697$\cdot10^{-3}$  &  -2.579$\cdot10^{-3}$ \\
    49  &  1.80  &  0.228  &  0.210  &  -1.688$\cdot10^{-3}$  &  -2.565$\cdot10^{-3}$ \\
    52  &  1.80  &  0.227  &  0.208  &  -1.679$\cdot10^{-3}$  &  -2.551$\cdot10^{-3}$ \\\hline
    \multicolumn{6}{|c|}{\sc A7 @ 0.6\,GHz}\\\hline
    {\smaller $T$} & {\sc afl} & {\sc fl-l} & {\sc fl-q} & {\sc rat-l} & {\sc rat-q} \\\hline
    50  &  2.99  &  0.359  &  0.327  &  -0.249$\cdot10^{-3}$  &  8.677$\cdot10^{-3}$ \\
    54  &  2.99  &  0.355  &  0.324  &  -0.247$\cdot10^{-3}$  &  8.615$\cdot10^{-3}$ \\
    58  &  2.99  &  0.352  &  0.321  &  -0.245$\cdot10^{-3}$  &  8.532$\cdot10^{-3}$ \\
    62  &  2.99  &  0.349  &  0.317  &  -0.243$\cdot10^{-3}$  &  8.430$\cdot10^{-3}$ \\\hline
    \end{tabular}
  \end{center}
  \label{table:fittingper}
\end{table}%
The maximum measured power fluctuation ({\sc afl}) due to varying temperature is shown to be between 1.21\% and 3\%.
The relative power fluctuation is computed as the relative transformed power fluctuation over the measured power fluctuations:
\begin{equation}
 \text{\sc fl} = \left. \frac{\max(P_r)-\min(P_r)}{\mathrm{median}(P_r)}\middle/\frac{\max(P_m)-\min(P_m)}{\mathrm{median}(P_m)}\right..
\end{equation}
It can be seen that the power fluctuations are diminished by a factor of about three to four in all cases.
This can also be visually verified in Figure~\ref{fig:waves}.
Moreover, the resulting transformation's distribution is quasi symmetric.
This is shown with the {\sc rat} metric in Table~\ref{table:fittingper}, which represents the departure of the median from the mean.
If the median differs significantly from the mean, then the distribution is not symmetric.
We see however that in all cases the median and mean are very close to each other, indicating that the transformation produces a symmetric distribution.
Now that the power is converted as if it were measured at a reference temperature, statistical methods have a more practical meaning.
Statistical methods applied directly to the measured power will produce estimates that are inflated based on the arbitrary distribution of the measured power samples.

We note that it is advisable to choose a reference temperature within the measured temperature range, preferably not too close to the extremities, to minimize transformation errors.

\section{Future Work}
\label{sec:futurework}

Enhancing the accuracy of the presented temperature/power models should be the main objective of future work.
A more profound study of the following observations, among others, is necessary.
What is the impact of \emph{temperature gradients} on the temperature/power behavior?
How do the \emph{initial conditions} of temperature and power affect the temperature/power behavior?
How is the \emph{accuracy of the temperature measurements} affected when the temperature sensor is not located at the temperature hotspots?
How can the \emph{temperature sensor accuracy} be improved?

The effects of temperature gradients and initial conditions can be assessed by controlling the testbed more precisely.
However, some parameters can only be controlled within certain limits, which poses great challenges for testbeds containing retail devices.
Also, the  accurate temperature sensors are not always available and often lack calibration facilities.

\begin{figure}[t!]
  \begin{center}
    \input{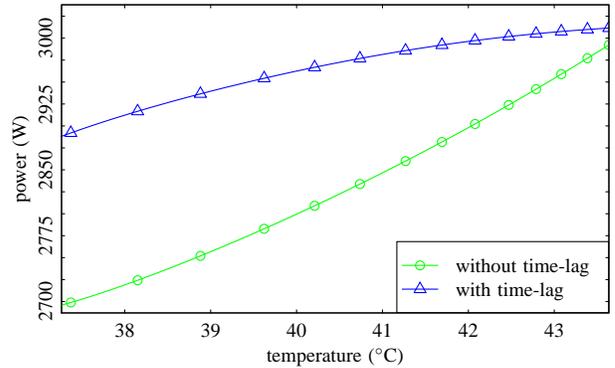}
  \end{center}
  \caption{Theoretical derivations predict monotonically increasing temperature/power curves, indicated by the green line.
	    Frequently, monotonically decreasing curves are measured (top line) as a result of heat propagation delays between temperature hotspots and the temperature sensor.
	    The bottom line was obtained by applying a transformation function to the top line.}
  \label{fig:expexample}
\end{figure}

Previously it was shown that the accuracy of temperature estimates depends, among other factors, upon the distance of the temperature sensor to the temperature hotspots in the processor~\cite{Kong:2012:RTM:2187671.2187675}.
Being located away from heat hotspots, the temperature sensor will not only provide an underestimation of the temperature, but the temperature measurements will also lag in time as a result of finite propagation delays, an issue referred to as the \emph{distant sensor syndrome}.
Such propagation delays may have some serious implications in extreme cases when the temperature sensors are off-board.
This will have an impact on the measured temperature/power relationship.
Indeed, there are generally two types of curves found in the literature, as shown in Figure~\ref{fig:expexample}.
First, the traces where the temperature/power curve bends downwards, as a result of the temperature lag over the power trace (blue line).
Examples of such measurements are the ones by Weissel and Bellosa~\cite{weissel04thermalmanagement}, and also our A9 measurements.
Measurements of the other type are traces where the temperature sensor is relatively close to the temperature hotspots; these also include results from simulations (green line).
Our traces, and the majority of the literature cited before, predict the upwards bending of the curve.

Let's assume that, for the downward bending curves, there exist a transformations function named $B(t)$.
The transformation function $B(t)$ transforms a temperature measurement of the distant temperature sensor into a measurement as it would have appeared to be at the hotspot: $T_\mathrm{CPU}(t) = B(t)\cdot T_\mathrm{sensor}(t)$.

A first order approximation to $B(t)$ could be constructed as follows.
Let's assume that our temperature sensor is located at a finite distance ($x=a$) in a finite one-dimensional slab.
A heat source with temperature $T_i$ is applied to the surface (at $x=0$ and $t=0$) of the semi-finite slab, with initial conditions $T_\infty$ for $x\rightarrow \infty$ at $t=0$.
The heat propagation in this case is described by Fourier's law of heat conduction.
The well-known solution for this setup is the error function $\theta(x,t) = \erf(x/\sqrt{4\alpha t})$, where $\alpha$ is the thermal diffusivity of the system and $\theta$ the relative temperature change~\cite{cengel2010heat}.
A more accurate approach would be to model the system in a two-dimensional space with additional thermal resistors to account for the multiple materials that the heat encounters when propagating from the temperature hotspot to the temperature sensor.
Assuming that the temperature at the \acs{CPU} hotspot increases exponentially, a first order approximation to the transformation function $B(t)$ can then be defined as
\begin{equation}
 B(t) = \frac{T_\mathrm{CPU}}{T_\mathrm{sensor}} = \frac{(T_\infty-T_i)~(1-\exp(-t/b))+T_i}{(T_\infty-T_i)~\erf(a/\sqrt{4\alpha t})+T_i},
\end{equation}
where $\alpha$, $a$ and $b$ are case-specific scalars.

Figure~\ref{fig:expexample} shows the effect of the transformation function on a temperature/power trace with constants $\alpha=4.125\cdot10^{-7}$, $a=8.25$ and $b=36.7$.
It is observed that the downward bending curve is indeed transformed into an upward bending curve as forecast by the theory.

The step response function $B(t)$ is however of limited practical value.
To reconstruct the temperature at the hotspot via a remote temperature sensor one also needs to know the time-dependent load on the system to assess the power consumption.
It is a question of acceptable overhead whether reconstructing the temperature is worthwhile pursuing via a transformation function like $B(t)$.

\section{Conclusion}
\label{sec:conclusion}

Via experimental data we have shown that the temperature/power relationship for some application processors shows a distinct exponential behavior, which is in line with theoretical foundations.
The exponential behavior is affected by, among other factors, temperature-dependent leakage currents, physical properties and the voltage regulator.

A practical model was presented to estimate the \acs{CPU} power consumption at a given temperature with arbitrary \acs{CPU} configurations.
We believe this model is useful for simulation purposes, although there is room to improve its accuracy.

We have also presented a real-life application where the effects of inflating temperature on power traces were removed.
We showed that the proposed technique can be quite effective.
The importance to know the whereabouts of the temperature onboard sensors is also pointed out.
The distance between temperature sensor and hotspot can significantly influence the shape of the temperature/power relationship.




\bibliographystyle{IEEEtran}
\bibliography{library}

\end{document}